\documentclass[12pt]{article}
\usepackage{amsmath,amsfonts,amssymb}
\usepackage{graphics}
\usepackage{lscape}
\usepackage{epsf}

\newcommand{\half}{{\textstyle{\frac{1}{2}}}}
\newcommand{\thalf}{{\textstyle{\frac{3}{2}}}}

 \newcommand{\be}{\begin{equation}}
\newcommand{\ee}{\end{equation}}
\newcommand{\bea}{\begin{eqnarray}}
\newcommand{\eea}{\end{eqnarray}}

\newcommand{\jn}{\Delta{J_{N}}}
\newcommand{\ji}{\Delta{J_{i}}}

\newcommand{\ai}{A_{i}}
\newcommand{\aik}{A_{i,k}}
\newcommand{\aikd}{A_{i,k}^{\dagger}}
\newcommand{\aid}{A_{i}^{\dagger}}
\newcommand{\jp}{J_{+}}
\newcommand{\jm}{J_{-}}
\newcommand{\s}[3]{\sum_{#1=#2}^{#3}}

\newcommand{\upa}{\uparrow}
\newcommand{\doa}{\downarrow}

\title{Quantum spin model fitting the Yule distribution of oligonucleotides in DNA}
 
\author{C. Minichini\thanks{E-mail: {\ttfamily minichini@na.infn.it}} \and A. Sciarrino\thanks{E-mail: {\ttfamily sciarrino@na.infn.it}}}

\date{}

\begin{document}

\begin{titlepage}

\maketitle
{\itshape{Dipartimento di Scienze Fisiche, Universit{\`a} di Napoli
``Federico II''
and I.N.F.N., Sezione di Napoli -
Complesso Universitario di Monte S. Angelo
Via Cintia, I-80126 Naples, Italy}}\\

\begin{abstract}
 A quantum spin chain is identified by the the labels of a  vector state of an 
 irreducible representation of $\mathcal{U}_{q \to 0}(sl_2)$. The intensity of 
 the one-spin flip is assumed to depend from the variation of the labels. The 
 rank ordered plot of the numerically computed, averaged in time, 
 transition probabilities  is  nicely fitted by a Yule distribution, which is the observed distribution
 of the ranked short oligonucleotides frequency in DNA
 \end{abstract}

\vfill

\begin{flushright}
preprint nr. DSF 29/2004
\end{flushright}

\end{titlepage}

\section{Introduction}

Quantum spin chains are extremely important tools to understand 
various physical situations and, in some cases, provide completely 
soluble models. An interesting field of applications of these models is 
the theory of molecular biological evolution.  Since 1986, when
Leuth$\ddot{a}$usser \cite{Leu86} put a correspondence between the Eigen model
  of evolution \cite{Eigen} and a two-dimensional Ising model, many 
  articles have been written representing biological systems as  
  spin models. Recently in \cite{BBW} it 
has been shown that the parallel mutation-selection model can be put in 
correspondence with the hamiltonian of an Ising quantum chain and in
\cite{Saakian} the Eigen model   has been 
mapped into the hamiltonian of one-dimensional quantum spin chains.
 In this approach the genetic sequence is specified by a
sequence of spin values $ \pm 1$.  DNA is build up a sequence of a four basis 
or nucleotides which are usually identified by their letter: C, G, T, A (T being replaced by
 U in RNA), C and U  (G and A) belonging to the purine family, 
 denoted by R (respectively to the pyrimidine family, denoted by Y).
 Therefore in the case of
 genome sequences each point in the 
 sequence
 should be identified by an element of a four letter alphabet. 
  As a simplification one identifies each 
 element according to the  purine or pyrimidine nature, reducing to a 
 binary value, see \cite{HWB} for a four-state quantum chain approach.
 The main aim of the works using this approach, see \cite{WBG,BW,HRW},
is to find, in different landscapes, the mean "fitness" and the 
"biological surplus", in the framework of biological population 
evolution.  As standard  assumption,   the strength 
of the mutation is assumed to depend from the Hamming distance 
between two sequences, i.e. the number of sites with different 
labels. Moreover usually it is assumed that the mutation matrix 
elements are vanishing for Hamming distances larger than 1, i.e. for 
more than one nucleotide changes. 
The Hamming distance assumption
is clearly
unrealistic in the domain of genetic mutations. 
On the other side Martindale and Konopka \cite{MK} have remarked that the 
ranked short (ranging from 3 to 10 
nucleotides) oligonucleotide frequencies, in both coding and non-coding region of DNA,
follow a Yule 
distribution. We recall that a Yule distribution is  given by $f = a \, {n^k}\, {b^n}$,
where $n$ is the rank and $a$, $ k < 0$ and $b$ are 3 real parameters.
  In this paper we propose a quantum 
 spin model   in which the intensity of the 
transition matrix depends in some way from the whole distribution of the 
nucleotides in the sequence. At present we assume that the transition matrix 
does not vanish only for total spin flip equal $\pm 1$, induced by the 
action of a single step operator, which generally is equivalent to one 
nucleotide change.
 The model, which can appear unphysical if applied to a 
 quantum spin chain, should be considered on the light of the previous 
 remarks on the application to the biological evolution and our aim 
 is to look for solutions which can reproduce the observed 
 oligonucleotides distribution.

%
%
%
%
%
%
%
%
%
%
 
\section{Mutations and Crystal basis}

A sequence of N ordered nucleotides, characterized only by the purine or 
pyrimidine character, can be represented as a vector belonging to the 
N-fold tensor product of the fundamental irreducible representation 
(irrep.) (labelled by $ J =1/2$) of $\mathcal{U}_{q \to 0}(sl_2)$ 
\cite{Kashi}, which is usually called crystal basis representation. This parametrization allows to represent, in a simple way,
the mutation of a  sequence as a transition between states, which
can be subjected to selection rules and  
whose strength  depends from the two concerned states. 
So we identify a N-nucleotidic sequence as a state
\be
 \mid \mathbf{J} \rangle = \mid J_3, J^N, \ldots,
          J^{i}, \ldots, J^2 \rangle
	\label{eq:def}
\ee 
where  $\; J^{N} \; $ labels the irrep. which the state belongs to,
$\; J_{3} \; $ is the value of the 3rd diagonal generator of
 $\mathcal{U}_{q \to 0}(sl_2)$ ($2J_{3} = n_{R} - n_{Y}$, $n_{x}$ 
being the number of $x$ elements in the sequence)
and $\; J^{i} \; $ ($ 2 \leq i \leq N - 1$) are $N-2$ labels needed
to remove the degeneracy of the irreps. in the tensor product  in 
order to 
completely identify the state and which can be seen as identifying the irrep.
 which the sequence truncated to the $i$-th element belongs to.
We introduce a scalar product, such that
\be
\langle \mathbf{J}\mid \mathbf{K}\rangle =
\left\{
\begin{array}{rl}
1 & \mbox{if } J_3 = K_3 \mbox{ and } J^i = K^i \; \forall i \\
0 & \mbox{otherwise}
\end{array}
\right.
\ee 
As an example, we can consider a trinucleotidic string
($N=3$) and label the eight different spin chains in the following way($\mid {J_3},{J^3},{J^2}\rangle$,
$R \equiv \half \equiv \upa, \; Y \equiv -\half \equiv \doa$): 
\begin{eqnarray*}
    \footnotesize
\upa\doa\doa &=& \mid -\half,\half,0 \rangle  \;\;\;\;
\upa\doa\upa = \mid \half,\half,0 \rangle \\
\doa\upa\doa &=& \mid -\half,\half,1 \rangle  \;\;\;\;
\upa\upa\doa = \mid \half,\half,1 \rangle \\
\doa\doa\doa &=& \mid -\thalf,\thalf,1 \rangle  \;\;\;\;
\doa\doa\upa = \mid -\half,\thalf,1 \rangle \\
\doa\upa\upa &=& \mid \half,\thalf,1 \rangle  \;\;\;\;\;\;\,
\upa\upa\upa = \mid \thalf,\thalf,1 \rangle.
\end{eqnarray*}     
Flipping the total spin by $\pm 1$ ($\Delta J_3= \pm 1 $) can induce a transition 
to a state belonging or not 
belonging to the irrep. of the original state. One can easily realize 
that  to identify a nucleotidic sequence as a state of 
an irrep. requires to fix the number of RY ``contracted couples" occurring 
in the considered sequence (contraction should be 
understood in the same sense of contraction of creation-annihilation 
operators in the Wick expansion). 
   Therefore flipping a spin implies or the creation   or the deletion 
  of 
   a RY contracted couple, corresponding respectively to a variation 
   of -1 o +1 of the value  of the $J^N$   and, possibly, of some others $J^i$
($2 \leq i \leq N-1$), or to leave unmodified the number of 
contracted couples (so that   $\Delta{J^{N}}=0$, but 
 $\Delta{J^{i}} \neq 0$ for some values of i). Note that alternatively one can
identify $1/2 \equiv (C,G)$ and $-1/2 \equiv (T,A)$. Below we give phenomenological
arguments for our choice.


\section{Transition operators}

Let us consider a N-nucleotidic string and classify the different transitions on the string labels $J_{3}, J^{N},\ldots,J^{2}$. We can distinguish different string configurations around the $i$-th position,
so that a single nucleotide mutation in $i$-th position can correspond to different variations in the string labels. We call \textit{left}
(\textit{right}) \textit{side free}  the nucleotides
on the left (right) of $i$-th position and not contracted    with
another one on the same side. 
Let $R_l$ be the initial (before mutation) number of the \textit{left side free} purines and $Y_r$
the initial number of the \textit{right side free} pyrimidines.
We want to count the total number of contracted $RY$ couples (before and after
mutation) in the
string, so we call $R_{in}$ ($R_{fi}$) the number, in the initial (final) state,
of $R$  preceding
some $Y$ and not contracted with any $Y$ on their side. In the same way, with $Y_{in}$ ($Y_{fi}$)
we refer to the number of $Y$ following some $R$ and not contracted with any $R$ on their side.
If a $R \rightarrow Y$ mutation ($\Delta{J_3}=-1$) occurs in $i$-th position, then $R_{in}=R_{fi}+1$
and $Y_{in}=Y_{fi}-1$, where $R_{in}=R_{l}+1$ and $Y_{in}=Y_{r}$. 
Now we can write  the transition part of the hamiltonian, for the 
different possible initial configuration of the string.
The transitions inducing operators  are built by means of $\jm,\ai,\aik$ and 
their adjoint operators, as below defined.   

\begin{itemize}

\item If $\mathbf{R_l}=\mathbf{Y_r}$ 
we distinguish two subcases:

\begin{enumerate}

\item $R_{l}=Y_{r}\neq{0}$
\be
{H_1}=\s{i}{2}{N-1}\s{k}{i+1}{N} \, \alpha_{1}^{ik} \,( {\aik\jm + \jp\aikd})
\label{eq:1}
\ee
\item $R_{l}=Y_{r}={0}$
\be
{H_2}=  \alpha_{2} \, (\jm+\jp) \label{eq:2}
\ee
\end{enumerate}
\item If $\mathbf{R_l}>\mathbf{Y_r}$
%
we distinguish two subcases:
\begin{enumerate}
\item $Y_{r}=0$ 
\be
{H_3}=\s{i}{2}{N} \, \alpha_{3}^{i} \, ({\ai\jm + \jp\aid}) \label{eq:3}
\ee
\item $Y_{r}\neq{0}$
\be
{H_4}=\s{i}{3}{N-1} \, \alpha_{4}^{i} \, ({\ai\jm + \jp\aid}) \label{eq:4}
\ee
\end{enumerate}
\item If $\mathbf{R_l}<\mathbf{Y_r}$
we distinguish two subcases:
\begin{enumerate}
\item $R_{l}=0$
 \be
{H_5}=\s{m}{2}{N} \, \alpha_{5}^{m} \,({\jm A_{m}^{\dagger} + A_{m}\jp}) 
\label{eq:5}
\ee
\item $R_{l}\neq0$
\be
{H_6}=\s{i}{2}{N-2}\s{k}{i+1}{N-1} \, \alpha_{6}^{ik} \,
({\aik \jm A_{k+1}^{\dagger} + \aikd A_{k+1} \jp})  \label{eq:6}
\ee
\end{enumerate}
\end{itemize} 
where  $\jp$ and $\jm$ are the \textit{step operators} defined by Kashiwara
\cite{Kashi}, acting on an
irreducible representation with highest weight $J^{N}$,  i.e. 
inducing 
the transitions $\ji=0, \; \forall i \neq N$,  
\bea
A_{i,k} \mid \mathbf{J} \rangle &=& \mid J_3, J^N,..,
        J^k, J^{k-1}-1,.., J^{i}-1, J^{i-1},.., J^2 \rangle 
 \nonumber \\     
& & (2 \leq i \leq N-1 \;\;\;\; i+1 \leq k \leq N)
\eea
\bea
A_i \mid \mathbf{J} \rangle &=& \mid J_3, J^{N}-1, \ldots,
    J^{i}-1, J^{i-1}, \ldots, J^2 \rangle
 \nonumber \\
& & (2 \leq i \leq N)
\eea
Therefore $\aikd$ is the operator which increase by 1 the value 
of $J^l$, for $k-1 \leq l \leq i$.
A few words to comment on the above equations. Let us consider a mutation $R \rightarrow Y$ which involve a transition
$\jn=-1$ (case $R_{l}>Y_{r}$); as of the considered transition also entails $\Delta{J_3}=-1$, we
have to apply the operator $\jm$, as well as the operator $\ai$. Of course, first we have to lower by
1 the value of $J_3$, then to modify $J^N$, otherwise the initial state may 
eventually be annihilated, even if the
transition is allowed (in the case $J^{N}-1<J_3$).
Likewise, in corrispondence of a transition $Y \rightarrow R$ ($\Delta{J_3}=+1$), first the change
$J^{N} \rightarrow J^{N}+1$ has to be maked, then $J_3 \rightarrow J_{3}+1$.
If we want to write a self-adjoint operator, we have to sum the operator which gives rise to the
transition $Y \rightarrow R$ with that one which leads to $R \rightarrow Y$, leaving the rest of the
string unmodified, so
\be
A_{i}J_- + {J_+}{A_{i}^{\dagger}}.
\ee
This operator leads to the mutation $Y \rightarrow R$ or $R \rightarrow Y$ for a nucleotide in
$i$-th position, in a string with $R_{l}>Y_{r}$.
If the mutation $R \rightarrow Y$ corresponds to a rising of $J^{N}$ (i.e. a transition with
$\jn=1,\Delta{J_3}=-1$, case $R_{l}<Y_{r}$), first $J^{N}$ has to be modified, then $J_3$;
therefore we write the self adjoint operator
\be
  {J_-}{A_{m}^{\dagger}} + {A_m}J_+
\ee
 The above operator gives rise to mutations $R \rightarrow Y$ and $Y \rightarrow R$
for a nucleotide in $i$-th position, preceding the $m$-th one, in the case $R_{l}=0, Y_{r}\neq{0}$.
%

%
%
%
%
%
%
%
%
%
%
%
%
%
%
%
%
%
Let us remark that eq.(\ref{eq:4}) is included in  
eq.(\ref{eq:3}), if the coupling constants $\alpha$ are assumed equal;
 in  eq.(\ref{eq:6}), only  the writing order for  $A_{k+1}$ (and its adjoint)
and $J_{\pm}$ has to be respected.
 Assuming now that the coupling constants do not depend on $i,k,m$, we can write 
 that transition hamiltonian $H_I$ as 
 \be
{H_I}=\mu_{1}({H_3}+{H_5})+\mu_{2}{H_1}+\mu_{3}{H_2}+\mu_{4}{H_6}
\ee
We let the fenomenology suggests us the scale of the values of the 
coupling constants of $H_I$. We want to write an interaction
term which makes the mutation in alternating purinic/pyrimidinic tracts less likely than
polipurinic or polipyrimidinic ones. We mean as a single nucleotide mutation in a polipurinic
(polipyrimidinic) tract, a mutation \emph{inside} a string with all nucleotides $R$ ($Y$), i.e. 
a highest (lowest) weight state. Such a transition corresponds to the 
selection rules
$\jn=-1$,$\Delta{J_3}=\pm1$, i.e. a transition generated by the action of $H_3$ and $H_5$.
In the interation term $H_I$, we give them a coupling constant smaller than the 
others terms.   
We introduce, for  $ \; \Delta J_{3} = \pm 1 \;$, only four
different mutation parameters $\;\mu_{i}\;$ ($i = 1,2,3,4$), with
${\mu_1}<{\mu_k} \; k>1$.
\begin{enumerate}
\item $\mu_{1}$ for mutations which change the irrep., $ \; \Delta {J^N} = \pm 
1 \; $, and include the spin flip inside an highest or lowest weight vector; 
\item $\mu_{2}$ for mutations which do not change the irrep., $ \; \Delta {J^N} = 0 
 \; $,  but modifies other values of $J^{k}$,  $\; \Delta J^{k} = \pm 1$ ($2 \leq k \leq N-1$);
\item $\mu_{3}$ for mutations which do not change the irrep., $ \; \Delta J^{N} = 0$,
      neither the other values of $J^{k}$, $\; \Delta J^{k} = 0$, ($2 \leq k \leq N-1$);  
\item $\mu_{4}$ for mutations which change the irrep., $ \; \Delta J = \pm 
1 \; $, but only in a string with $0 \neq R_{l} < Y_{r}$.
\end{enumerate}
 We do not introduce another parameter, for mutations
generated by $H_4$, i.e. $i$-th nucleotide mutation in a string with
$R_{l} > Y_{r} \neq 0$, to not distinguish, in a polipurinic string, between 
a mutation in $2$-th position and another one inside the string.
    
%
%
Let us emphasize once more that the  proposed model  
takes into account, at least partially, the effects on the transition in the $i$-th site of the 
distribution of all the spins.


\section{Results}

%
%
The total hamiltonian of the model will be written as $H = H_{0}+H_I$,
where $ H_{0}$ is the diagonal part in the choosen basis and,  
in the following, is assumed to be  $H_{0} =  \mu_{0}J_{3}$. 
In order to evaluate the probabilities of transition, we cannot analytically study the time evolution of an initial state, 
 representing a given spin chain,  as ruled by hamiltonian $H$. So we 
look for a numerical solution, with a suitable choice of the value of 
the parameters, for N = 3,4,6.
Before solving numerically the model, it may be useful to point out
 explicitly its main features. $H_0$ is the hamiltonian of a spin chain
 in presence of an uniform, constant, magnetic field (in the biological
 system it is the "fitness" assumed proportional to total purine
 surplus). $H_I$
 describes an interaction on the $i$-th spin neither depending on the
 position nor on the nature of the closest neighbours. Indeed it
 depends on the "ordered" spin orientation surplus on the left and on
 the right of the $i$-th position. Should it not depend on the order, it
 may be considered as a mean-field like effect. Moreover $\Delta J_3 =
 \pm 1$ transitions   are allowed, which can be considered or as the
 flip of a spin combined with an exchange of the two, oppositely
 oriented, previous or following spins or as the collective flip of
 particular three spin systems, containing a two spin system with
opposite spin orientations (see example below).
Biologically, the transition depends in some way on the  "ordered"
purine surplus on the left and on the right of the mutant position.
For example, the matrix form of $H$, on the above basis (for N = 3) is the following
one, up to a multiplicative dimensional factor $\mu_{0}$
\begin{equation}
\label{matrixModel}
H = 
\left(
\begin{array}{cccccccc}
-1 & \delta & 0 & \gamma & \epsilon & 0 & \epsilon & 0 \\
\delta & 1 & 0 & 0 & 0 & \epsilon & 0 & \epsilon \\
0 & 0 & -1 & \delta & \epsilon & 0 & \epsilon & 0 \\
\gamma & 0 & \delta & 1 & 0 & \epsilon & 0 & \epsilon \\
\epsilon & 0 & \epsilon & 0 & -3 & \delta & 0 & 0 \\
0 & \epsilon & 0 & \epsilon & \delta & -1 & \delta & 0 \\
\epsilon & 0 & \epsilon & 0 & 0 & \delta & 1 & \delta  \\
0 & \epsilon & 0 & \epsilon & 0 & 0 & \delta & 3
\end{array}
\right)  
\end{equation}
Note that the above hamiltonian depends only on three coupling constants
due to the very short lenght of the chain. For $N \ge 4$ the 4th coupling constant
(denoted in the following by $\eta$) will appear.
Let us emphasize that the hamiltonian (\ref{matrixModel}) does not only
connect states at unitary Hamming distance. As an example, we look at the
transitions allowed from $\mid \half,\half,0\rangle$ ($\upa\doa\upa$) and from
$\mid -\half,\half,0\rangle$ ($\upa\doa\doa$)
\begin{eqnarray*}
\upa\doa\upa \longrightarrow
\left\{
\begin{array}{c}
\upa\upa\upa \\
\doa\doa\upa \\
\upa\doa\doa
\end{array}
\right.& \qquad \upa\doa\doa \longrightarrow
\left\{
\begin{array}{c}
\doa\upa\upa \\
\doa\doa\doa \\
\upa\upa\doa \\
\upa\doa\upa
\end{array}
\right.
\end{eqnarray*}
%

\begin{figure}[t]
\begin{center}
\framebox{\epsfxsize=0.70\textwidth
\epsffile{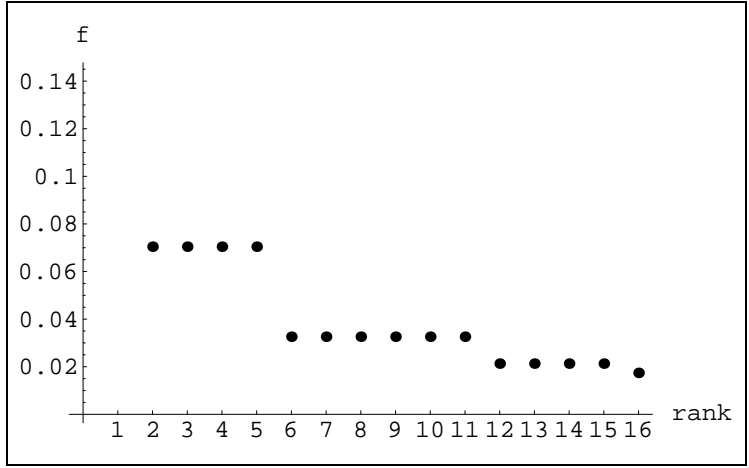}}
\end{center}
\caption{Rank ordered distribution of time-averaged transition probability from the initial
status $\upa\upa\doa$ for a dynamics generated by Hamming hamiltonian, with 
$\beta=0.5$. The first point is out of the graph.}
\label{stepH8}
\end{figure}

\begin{figure}[b]
\begin{center}
\framebox{\epsfxsize=0.70\textwidth
\epsffile{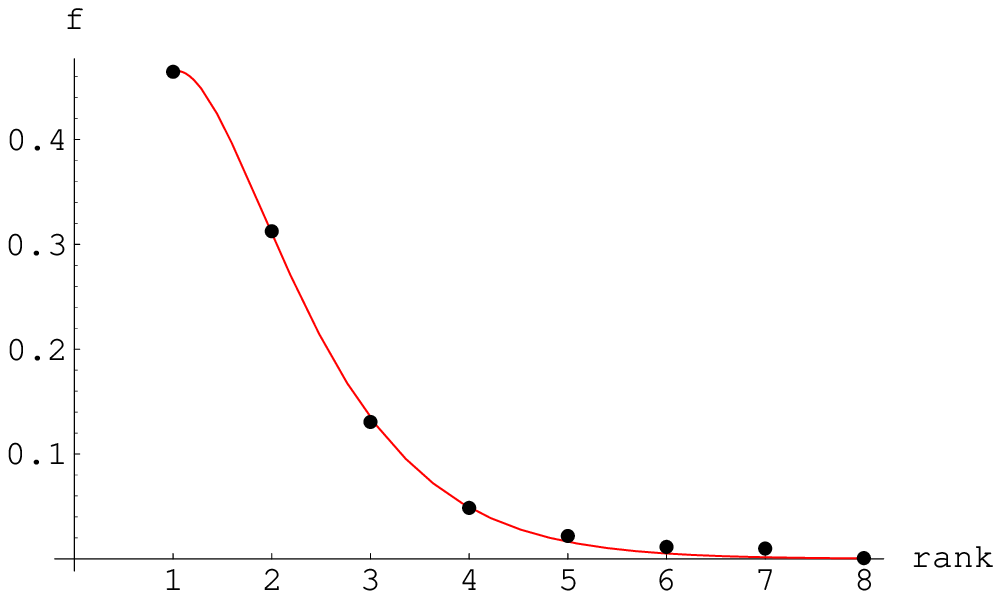}}
\end{center}
\caption{Rank ordered distribution of time-averaged transition probability $f$ from the initial
status $\upa\upa\doa$ for a dynamics generated by $H$ with $\epsilon=0.1,\;\gamma=\delta=0.3$.
The distribution was fitted by a Yule function (continuous line) $f=a{R^k}{b^R}$ ($R$ is the
rank).The parameters was estimated as $a=1.96,\,b=0.24,\,k=-1.49$.}
\label{YuleH8}
\end{figure}

%
 The transition probability between two states shows the typical
quantum-mechanical rapidly oscillating  behaviour in time.
  We define a time-averaged transition probability ($ i \rightarrow f$)
  \be
  <p_{if}> = \frac{1}{T} \, \int_{0}^{T} \, p_{if}(t) \, dt
  \label{eq:ta}
  \ee
where the value of $T$ will be numerically fixed to a value, such 
that the r.h.s. of eq.(\ref{eq:ta}) is stable. For lack of space, we do not explicitly write the 
hamiltonian which only induces transitions between chains at
Hamming distance equal to one, with coupling constant $\beta$. Notice howevere that such a (Hamming) hamiltonian is 
not obtained by  eq.(\ref{matrixModel}) putting  $ \delta  = \gamma = 
\epsilon = \beta$. 
If we order (in a decreasing way) the average transition probability from an 
initial state to every
other state, we obtain, using the hamiltonian with Hamming distance, a rank
ordered distribution of transition probability like that in fig.\ref{stepH8} for $N=4$. Its 
shape does not depend by the choice of initial state or by the coupling
constant $\beta$ value.
We always get the same structure, for models with transition probability
only depending on Hamming distances. So the rank-ordered distribution
of the average transition probability shows a plateaux structure: every
plateaux contains spin sequences at the same Hamming distance from the initial one.
In the case of the model which we propose here, i.e. the hamiltonian in
(\ref{matrixModel}), the distribution
of rank ordered average transition probability does not show a 
plateaux structure,
but its shape is well fitted by a Yule distribution (fig.\ref{YuleH8}), like the rank ordered 
distribution of oligonucleotidic frequency in the strings of nucleic acids
\cite{MK}.

%

\begin{figure}[t]
\begin{center}
\framebox{\epsfxsize=0.69\textwidth
\epsffile{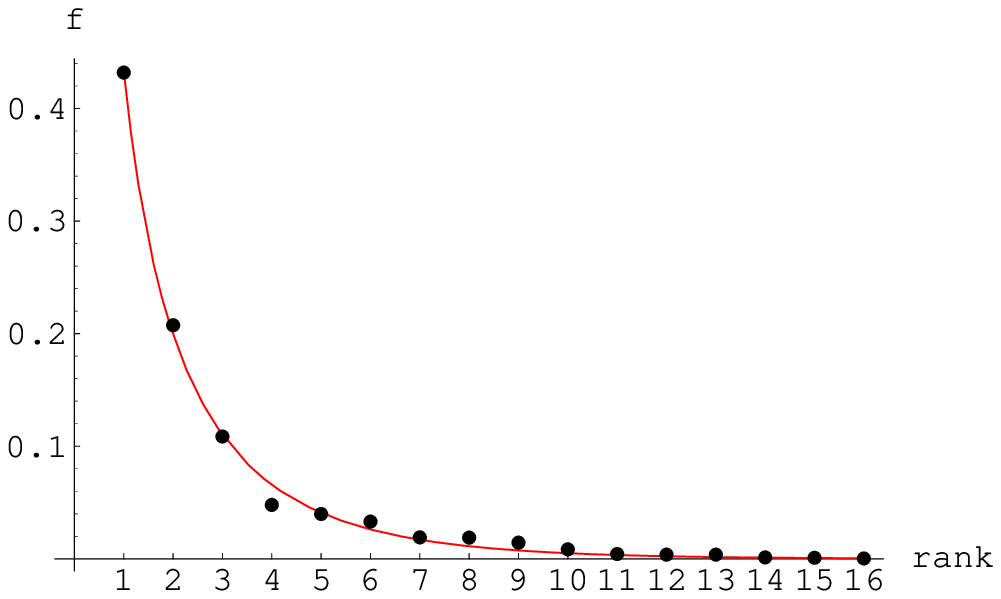}}
\end{center}
\caption{Rank ordered distribution of time averaged transition probability $f$ from
the initial status $\doa\doa\upa\doa$ for a dynamics generated by $H$, with
$\epsilon=0.1,\;\gamma=\delta=\eta=0.5$. The distribution was fitted by a Yule
function (continuous line) $f=a{R^k}{b^R}$ ($R$ is the rank). The parameters was estimated as $a=0.60,\,b=0.72\,k=-0.64$.} 
\label{N=4}
\end{figure}

\begin{figure}[b]
\begin{center}
\framebox{\epsfxsize=0.69\textwidth
\epsffile{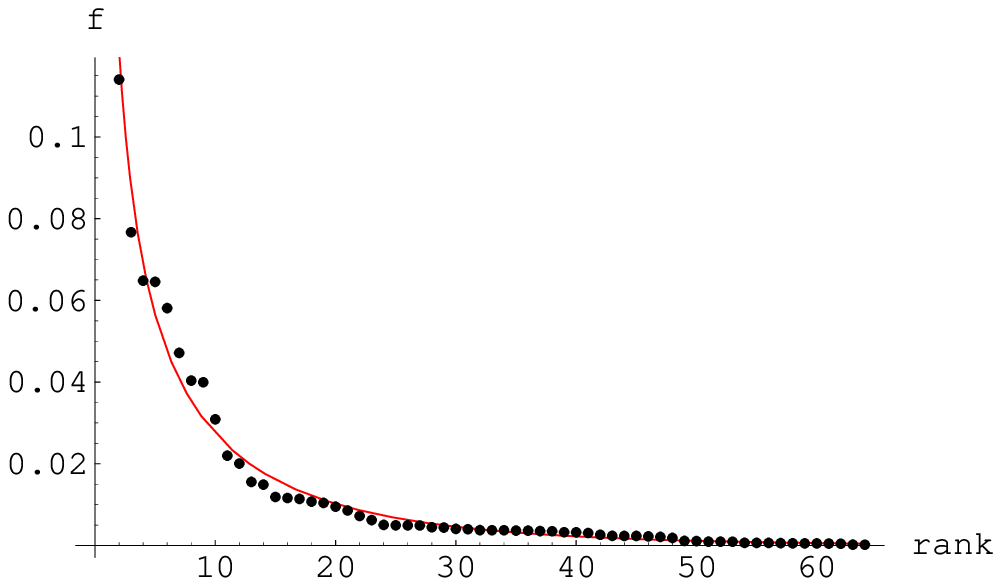}}
\end{center}
\caption{Rank ordered distribution of time averaged transition probability $f$ from
the initial status $\upa\doa\upa\doa\upa\doa$ for a dynamics generated by $H$, with
$\epsilon=0.1,\;\gamma=\delta=\eta=0.5$. The distribution was fitted by a Yule
function (continuous line) $f=a{R^k}{b^R}$ ($R$ is the rank). The parameters was estimated as $a=0.21,\,b=0.95\,k=-0.66$.} 
\label{N=6}
\end{figure}

%
Let us observe that we obtain a Yule distribution (and not a plateaux structure)
even if all parameters in (\ref{matrixModel}) are tuned at the same 
value, which means that the distribution is the outcome of the model 
and not of the choice of the values of the coupling constants.
Analogous resultes are gained for $N=4$ (fig.\ref{N=4}) and $N=6$ 
(fig.\ref{N=6}), the state labelled by 1 being the initial one.

%
%

As final remarks we point out that:

i) our model is not equivalent to a model where
  the intensity  depends on the site  undergoing the 
  transition, or from the nature of the closest neighbours or the 
  number of the $R$ and $Y$ labels of the sequence; indeed essentially 
  the intensity depends on distribution in the sequence of 
  the $R$ and $Y$;
  
ii) the ranked distribution of the 
probabilities, not averaged in time, computed for several values of 
the time follows generally a Yule distribution law;
for the highest value of $N$, the distribution is equally well 
fitted by a Zipf law, but not for the lowest values of $N$, in 
agreement with the remark of \cite{MK}.

In conclusion we are far from claiming that our simple quantum 
mechanical model explains the observed oligonucleotide distribution 
for several obious reasons. Apart from the extremely simplifying 
assumption made (initial state as a pure state) and from the fact that 
we are really comparing different data, the experimental ones being 
derived from  the splitting in short oligonulceotide sequences of 
several far longer sequences, we are  relating
quantities computed in a quantum world to classical observables.
In any way we believe that 
the proposed model exhibits intriguing  and interesting features, 
hinting in the right direction, which are worthwhile to be further 
investigated, in particular either to make more clear the connection
quantum model-classical quantities either to reformulate our 
transition matrix in order to be inserted in a classical master equation.
It is worthwhile to remark that we are trying to compare theoretical 
results, deriving from a simple model (depending only on 4 parameters for any N), to really observed data, coming 
from the extremely complex biological world.
More details and further developments will be presented in a longer 
paper.

\newpage

\end{document}